# Contrast-Free Ultrasound Microvascular Imaging via Radiality and Similarity Weighting

Jingyi Yin, Jingke Zhang, Lijie Huang, U-Wai Lok, Ryan M DeRuiter, Kaipeng Ji, Yanzhe Zhao, Kate M. Knoll, Kendra E. Petersen, Tao Wu, Xiang-yang Zhu, James D Krier, Kathryn A. Robinson, Lilach O Lerman, Andrew J. Bentall, Shigao Chen, and Chengwu Huang

(*Corresponding authors: Shigao Chen and Chengwu Huang*)

J. Yin, J. Zhang, L. Huang, U.-W. Lok, R. M. DeRuiter, K. Ji, Y. Zhao, K. M. Knoll, K. E. Petersen, K. A. Robinson, S. Chen, and C. Huang are with the Department of Radiology, Mayo Clinic College of Medicine and Science, Rochester, MN 55905 USA (e-mail: Chen.Shigao@mayo.edu, Huang.Chengwu@mayo.edu).

T. Wu is with the Department of Medical Ultrasonics, The Third Affiliated Hospital of Sun Yat-Sen University, Guangzhou 510080 China, and the Department of Radiology, Mayo Clinic College of Medicine and Science, Rochester, MN 55905 USA.

X.-Y. Zhu, J. D. Krier, L. O. Lerman, and A. J. Bentall are with the Division of Nephrology and Hypertension, Mayo Clinic, Rochester, MN 55905 USA.

# Abstract

Microvascular imaging has advanced significantly with ultrafast data acquisition and improved clutter filtering, enhancing the sensitivity of power Doppler imaging to small vessels. However, the image quality remains limited by spatial resolution and elevated background noise, both of which impede visualization and accurate quantification. To address these limitations, this study proposes a high-resolution cross-correlation Power Doppler (HR-XPD) method that integrates spatial radiality weighting with Doppler signal coherence analysis, thereby enhancing spatial resolution while suppressing artifacts and background noise. Quantitative evaluations in simulation and *in vivo* experiments on healthy human liver, transplanted human kidney, and pig kidney demonstrated that HR-XPD significantly improves microvascular resolvability and contrast compared to conventional PD. *In vivo* results showed up to a 2 to 3-fold enhancement in spatial resolution and an increase in contrast by up to 20 dB. High-resolution vascular details were clearly depicted within a short acquisition time of only 0.3 s-1.2 s without the use of contrast agents. These findings indicate that HR-XPD provides an effective, contrast-free, and high-resolution microvascular imaging approach with broad applicability in both preclinical and clinical research.

# Introduction

Power Doppler (PD) imaging is a well-established and widely implemented technique for blood flow analysis and quantification, playing an indispensable role in clinical diagnosis [1, 2]. In the past decade, advancements in high-frame-rate plane wave imaging and tissue clutter filtering, particularly using singular value decomposition (SVD)-based methods, have significantly improved the sensitivity of PD imaging to slow blood flow [3-14], substantially broadening the scope of microvascular imaging [6, 8, 15, 16]. However, like all conventional ultrasound modalities, it is fundamentally constrained by the diffraction limit, which restricts the achievable spatial resolution [17]. Higher resolution would enhance visualization and accurate monitoring the blood flow in small vessels, and reliable quantification of vascular characteristics, which can be beneficial for future clinical diagnosis. Increasing evidence suggests that abnormalities in microvasculature can serve as early indicators of diseases, such as neurological disorders [18, 19], inflammatory bowel disease [20], and tumor diagnosis [21]. This highlights the need for imaging techniques with both high spatial and temporal resolution.

In recent years, there has been growing interest in high-resolution ultrasound imaging methods that utilize backscattered signals from native red blood cells (RBCs) to further enhance the quality of microvascular imaging. Efforts have focused on improving spatial resolution through spectral decomposition [22], deconvolution [23], harmonic imaging [24-27], deep learning [28], and speckle-based localization [29-31]. Despite these promising results, these methods remain technically challenged by the weaker and denser signal nature of RBCs compared to MBs [28], as well as the low signal-to-noise ratio (SNR), especially in deep tissue imaging [5].

Although plane wave compounding increases sensitivity to microvascular flow by collecting large temporal ensembles in a short time, it has less penetration and thus more susceptible to noise [9]. Moreover, the low penetration depth of unfocused plane waves exacerbates SNR degradation in deep organs [9]. For *in vivo* human imaging, this is particularly problematic because the complex, spatially varying tissue structures and scanning settings could degrade signal quality and generate artifacts that significantly limit the visibility of small vessels in deep regions.

To address these challenges, several approaches explore the coherence of blood signals in angular, temporal or spatial/channel domain, therefore enhancing contrast effectively. For instance, coherent flow power Doppler (CFPD) exploits the spatial coherence of blood signals to improve flow detection and suppress uncorrelated noise [32-34]. A non-normalized coherence metric has been proposed to preserve the linear relationship between image intensity and blood echo magnitude [35]. Acoustic subaperture processing (ASAP), which divides channel data into non-overlapping subgroups, enhances contrast by suppressing uncorrelated noise [36, 37]. Pialot *et al.* exploited coherence information between channels to identify noisy pixels [38]. In addition, higher-lag autocorrelations have been employed to reduce Doppler noise floors [39]. Huang *et al.* introduced the spatiotemporal nonlocal means filtering (ST-NLM) for clutter-filtered blood flow radio-frequency [40]. Jakovljevic *et al.* estimated angular coherence from the beamsummed signals to suppresses noise and motion artifacts in ultrafast data [41]. Huang *et al.* proposed correlating flow signals across transmit angles to suppress incoherent clutter and enhance visualization of slow flow [5, 42]. A nonlinear compounding method, denoted as frame-multiply-and-sum (FMAS), estimate signal coherence using autocorrelation function over plane-wave angle frames [43]. Delay-multiply-and-sum (DMAS) beamforming exploits signal coherence of the echo matrix in the dimension of transmit angle[44]. A united spatial–angular adaptive scaling Wiener postfilter was also developed to suppress noise [45]. While these approaches significantly improve contrast, there remains a pressing need for methods that can simultaneously enhance contrast and resolution.

In this work, we propose a high-resolution cross-correlation power Doppler (HR-XPD) strategy. The core idea is to enhance vascular image quality by jointly leveraging spatial features and spatial-temporal coherence of ultrasound signals. By combining similarity mapping from two angular Doppler subsets with radiality-based spatial weighting, HR-XPD achieves higher vascular resolution and stronger background suppression than conventional Doppler approaches.

The HR-XPD framework is validated in simulations, *in vivo* human liver, and kidney transplant, as well as in a pig kidney. We also evaluate its performance under varying noise conditions, and benchmark against conventional PD image similarity-map-weighted PD image (SW-XPD) image, and high-resolution power Doppler (HR-PD) image.

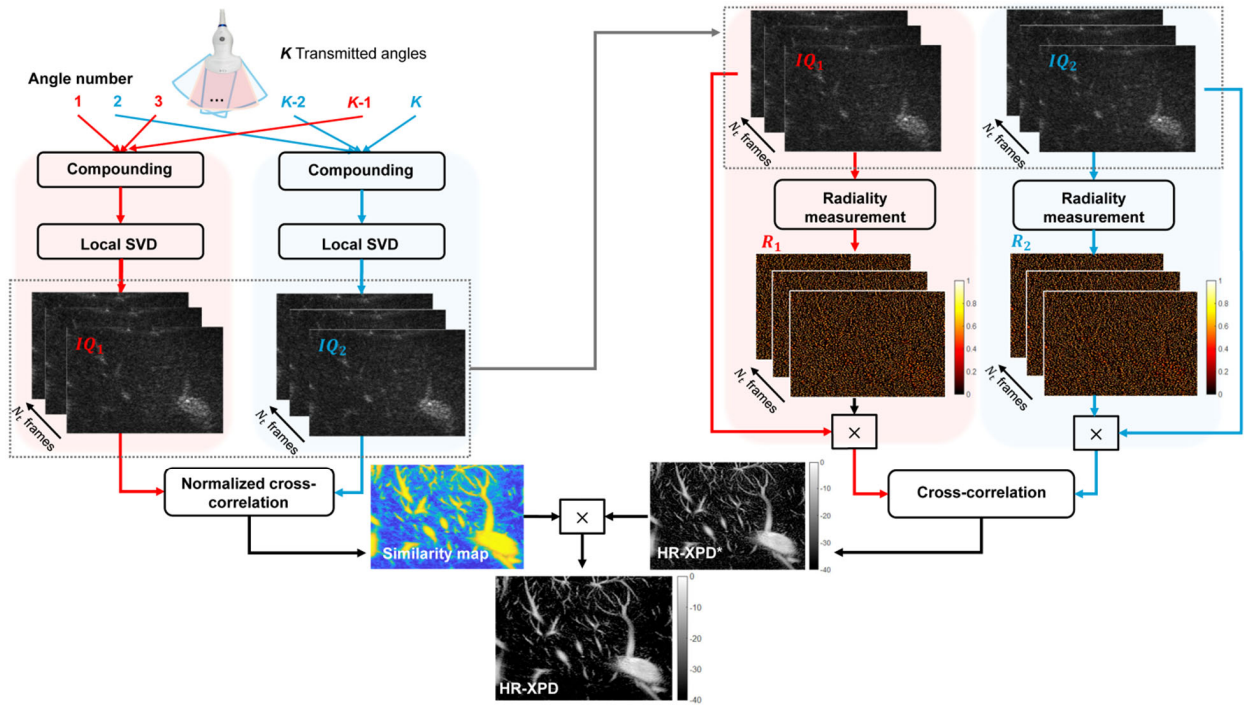

**Fig. 1. Schematic diagram of the proposed HR-XPD.** A total of $K$ transmissions at different steering angles are divided into two interleaved subgroups. Each subgroup undergoes compounding followed by local-SVD clutter filtering to obtain two sets of clutter-filtered IQ data. Normalized cross-correlation between the two subsets ($IQ_1$ and $IQ_2$) produces a similarity map that highlights coherent flow signals. In parallel, radiality maps ($R_1$ and $R_2$) are computed from the compounded subsets $IQ_1$ and $IQ_2$, and applied back to their respective IQ data, followed by cross correlation to form a HR-XPD* image with improved spatial resolution. The final HR-XPD image is obtained by multiplying the HR-XPD* image with the similarity map, with further suppression of background noise.

# Materials and Methods

## Principle of HR-XPD

The processing pipeline is illustrated in Fig. 1. The proposed method utilizes ultrafast ultrasound imaging, where multiple unfocused waves are typically transmitted sequentially with different steering angles, and their corresponding beamformed data are coherently summed to generate a compounded ultrasound image [6]. Instead of summing all the angles, the proposed method divides the steered transmissions into two groups, followed by compounding within each group to generate two compounded datasets, as shown in Fig. 1 [5].

We assume a total of $N_t$ frames were acquired (Fig. 1). During each frame acquisition, a total number of $K$ plane waves were transmitted at different steering angles. Odd-numbered angles (1, 3, …, $K-1$) were beamformed and compounded to form the first dataset, and even-numbered angles (2, 4, …, $K$) were beamformed and compounded to form the second dataset. Given the complex spatial variability of tissue clutter and background noise in large field-of-view human imaging, a block-wise adaptive SVD filtering strategy was adopted to extract the blood flow signals from the two compounded datasets [5, 9, 46]. Based on previous studies [5, 9], the local block-wise SVD approach outperformed global SVD methods, demonstrating significantly improved clutter suppression and blood signal extraction for *in vivo* kidney and liver imaging [5, 9]. The resulting clutter-filtered datasets were denoted as $IQ_1 = \{IQ_1(x, z, t) \,|\, t = 1, 2, \ldots, N_t\}$ and $IQ_2 = \{IQ_2(x, z, t) \,|\, t = 1, 2, \ldots, N_t\}$, as shown in Fig. 1.

Previous studies using contrast-enhanced ultrasound (CEUS) have demonstrated that the evaluation of the local gradient convergence degree (radiality) can serve as a weighting factor to reduce the size of the microbubble point spread function (PSF), thereby enabling non-localization-based super-resolution ultrasound (SRUS) imaging [47-49]. Building on this concept, we quantified local radiality from the envelopes of $IQ_1$ and $IQ_2$, and employed it as a weighting factor to enhance pixel intensities at PSF centers in the spatial domain without need of contrast agent (Fig. 1). Specifically, for each pixel coordinate $q = (x_q, z_q)$, the radiality $G_q$ at pixel $q$ was defined as a measure of the symmetry of the local gradient distribution [47, 48, 50].

The calculation of $G_q$ proceeded as follows. First, a circular annulus centered at pixel $q$ with an inner radius $r$ ($r = 0.5\ pixel$) around pixel $q$ was defined, and 8 sampling points $\{(x_i, z_i)|(i = 1, 2, \ldots 8)\}$ were evenly distributed along its circumference. Second, the spatial gradient vectors $\nabla I(x_i, z_i)$ were calculated at all sampling points. Third, for each sampling point, a radial vector $\mathbf{r}_i = (x_i - x_q, z_i - z_q)$ was computed, and the alignment between $\nabla I(x_i, z_i)$ and $\mathbf{r}_i$ was quantified using the normalized dot product. The resulting radiality value at pixel $q$ was given by:

$$G_q = \frac{1}{8}\sum_{i=1}^{8} \frac{\nabla I(x_i,z_i) \cdot \mathbf{r}_i}{\|\nabla I(x_i,z_i)\|\|\mathbf{r}_i\|} \tag{1}$$

This procedure was repeated for all pixels to produce radiality maps for both $IQ_1$ and $IQ_2$ across

$N_t$ frames, denoted as $R_1 = \{R_1(x,z,t) | t = 1, 2, ..., N_t\}$ and $R_2 = \{R_2(x,z,t) | t = 1, 2, ..., N_t\}$, respectively.

However, direct application of radiality weighting in contrast-free imaging is more challenging because radiality is highly sensitive to noise and background interference [48, 50]. In noisy conditions, the radiality map becomes unstable and tends to preserve artifacts, which may degrade image quality. To address this, we incorporated Doppler signal coherence analysis. Since noise is random and uncorrelated between angular data subsets, whereas blood flow signals remain relatively coherent [5], the radiality maps $R_1$ and $R_2$ ($t = 1,2,...,N_t$) were used to spatially weight $IQ_1$ and $IQ_2$ separately, leading to weighted in-phase and quadrature (IQ) data $IQ_{w1}$ and $IQ_{w2}$, followed by cross-correlation operation:

$$\text{HR-XPD*} = \left|\sum_{t=1}^{N_t} IQ_{w1} \cdot IQ_{w2}^*\right| \quad (2)$$

where $IQ_{w2}^*$ denotes the conjugate operation of $IQ_{w2}$. Before weighting, $R_1$ and $R_2$ are log-compressed and normalized to compensate for the change of dynamic range in radiality estimation. When assuming the clutter filtered IQ data consists of blood flow signal and additive noises, the cross-terms in Eq. 2 associated with noise interference are suppressed, while the coherent blood flow signal is preserved with high resolution (Details in [5]).

To further suppress strong background interference, we then calculated similarity map $M_0(x,z)$, which is denoted as the normalized cross-correlation between $IQ_1$ and $IQ_2$ [42]:

$$M_0(x,z) = \frac{\left|\sum_{t=1}^{N_t} IQ_1 \cdot IQ_2^*\right|}{\sqrt{(\sum_{t=1}^{N_t} |IQ_1|^2)(\sum_{t=1}^{N_t} |IQ_2|^2)}} \quad (3)$$

In this way, the coherent blood flow signals across subsets are maximized in $M_0(x,z)$, while incoherent noise and artifacts can be minimized. $M_0(x,z)$ is therefore used as a weighting map to enhance microvascular visualization, yielding the HR-XPD image:

$$\text{HR-XPD} = 10 \cdot log_{10}[M_0(x,z) \cdot \text{HR-XPD*}(x,z)] \quad (4)$$

For comparison, HR-PD images were obtained by applying radiality weighting to each subset individually:

$$\text{HR-PD} = 10 \cdot log_{10}\left[\sum_{t=1}^{N_t} |IQ_1 \cdot R_1 + IQ_2 \cdot R_2|^2\right] \quad (5)$$

A high-contrast power Doppler image, denoted as SW-XPD, was also generated by directly correlating the two clutter-filtered datasets and weighting similarity map [5, 42]:

$$\text{SW-XPD} = 10 \cdot log_{10}[\ \left|\sum_{t=1}^{N_t} IQ_1 \cdot IQ_2^*\right| \cdot M_0(x,z)] \quad (6)$$

## Simulation

In this study, two crossing tubes with diameter of 1 mm were simulated using Field II software [51-53]. Point scatterers were initially distributed randomly within the pre-set tube area with a density of 20 scatterers per mm3. Their positions were updated over time according to predefined uniform flow velocities (30 mm/s), under a plane-wave data acquisition at a frame rate of 500 Hz. Simulation parameters were listed in Table 1.

To evaluate the performance of the proposed method under varying noise levels, five different SNR settings (-2, 3, 8, 12, 18 dB) were simulated. Gaussian random noises were generated and added to the beamformed IQ subsets $IQ_1$ and $IQ_2$, respectively.

TABLE I
SIMULATION SETTINGS

|  | Value |
|---|---|
| Transmit center frequency (MHz) | 5.21 |
| Number of elements | 128 |
| Pitch (mm) | 0.3 |
| Transmission type | Plane wave |
| Steering Angles | −7.5°, −3.75°, 0°, 3.75°, 7.5° |
| Post-compounded frame rate (Hz) | 500 |
| Total number of frames | 300 |
| Number of pulse cycles | 2 |
| Pixel size of IQ (mm × mm, axial × lateral) | 0.04 × 0.15 |

## *In vivo* human data

The *in vivo* human study was approved by the Institutional Review Board (IRB) of Mayo Clinic, written informed consent was obtained from each participant. A Vantage 256 ultrasound system (Verasonics Inc., Kirkland, WA, USA) and a GE 9L linear array transducer (GE Healthcare, Wauwatosa, WI, USA) were utilized for transplanted kidney data acquisition. Ultrasound imaging of a healthy human liver was performed using the same ultrasound system and a C1-6-D probe (GE Healthcare, Wauwatosa, WI, USA). The volunteer was instructed to hold the breath during each data acquisition session to minimize tissue motion. The beamformed IQ data were obtained from the Verasonics ultrasound system. The parameter settings are shown in Supplementary Table S1.

### Pig kidney

The pig study was approved by the Mayo Clinic Institutional Animal Care and Use Committee (IACUC). In this study, we used a pig model of chronic kidney disease (CKD), and imaging was performed on the contralateral kidney. Detailed of pig study has been provided in [54]. The imaging parameter settings are shown in Supplementary Table S1.

### Quantitative Evaluation

Quantitative indexes, including contrast ratio (CR), full width at half-maximum (FWHM), resolution improvement ratio, background intensity, and peak-to-valley level (PVL) were computed to assess PD images.

The CR is calculated as follows [55]:

$$CR = \frac{|\bar{S}_{vessel} - \bar{S}_{background}|}{\sqrt{\bar{S}_{vessel}^2 + \bar{S}_{background}^2}} \tag{7}$$

where $\bar{S}_{vessel}$ and $\bar{S}_{background}$ are the mean intensity of the vessel region (or vessel peak) and background area. The CR ranges from 0 to 1, with higher values indicating better vessel-to-background contrast.

The FWHM was measured from the cross-sectional profile of local vessels, defined as the −3 dB width of the PD profile [34].

The resolution improvement ratio was computed as the ratio of FWHM values between different methods for the same vessel cross-section. Specifically, we denoted the ratio as $FWHM_{PD}/FWHM_{HR-XPD}$ and $FWHM_{SW-XPD}/FWHM_{HR-XPD}$.

The background intensity is calculated as follows [34]:

$$Background\ intensity = 10 \cdot log_{10}(S_{background})[dB] \tag{8}$$

where $S_{background}$ denote the power Doppler values of background pixels.

PVL is calculated as:

$$PVL = 10 \cdot log_{10}(\frac{\bar{S}_{peaks}}{\bar{S}_{valley}})[dB] \tag{9}$$

where $\bar{S}_{peaks}$ is the mean peak intensity of the two neighboring vessels along the sampling line and $S_{valley}$ is the valley intensity between two peaks [34]. PVL < 3 dB indicates that the valley intensity exceeds half the peak intensity, and the two nearby vessels are considered unresolved.

### Statistics

FWHM measurements, CR, and resolution improvement ratios were compared using one-way analysis of variance (RM one-way ANOVA). Post-hoc pairwise comparisons were performed with Geisser–Greenhouse correction. Statistical significance was set at p < 0.05. All analyses were conducted using GraphPad Prism 9.0 (GraphPad Software, San Diego, CA, USA).

## Results

### Simulation results

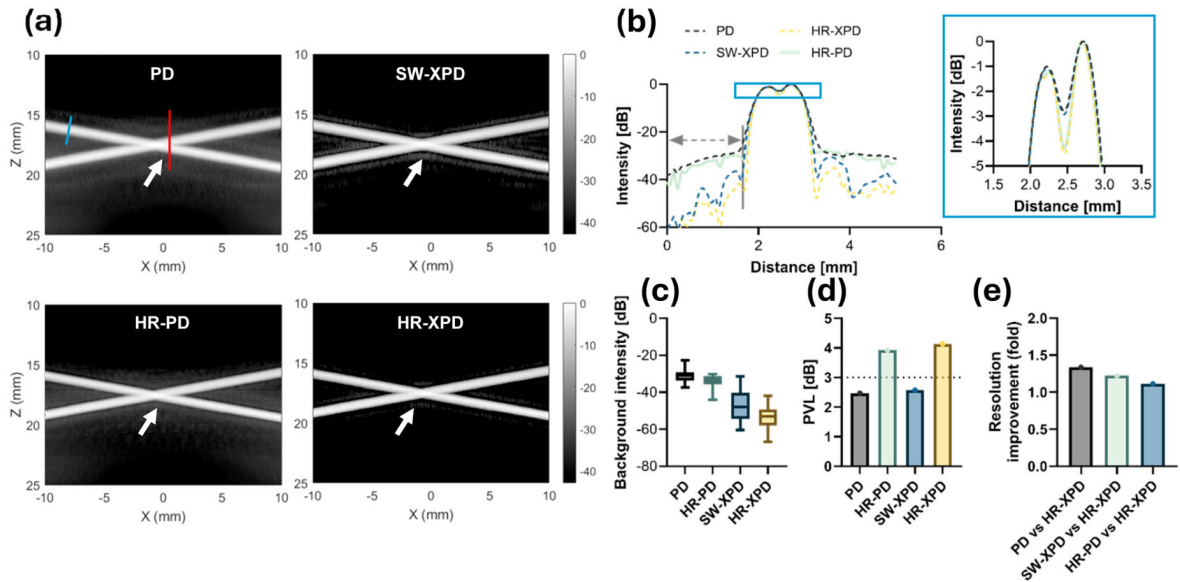

**Fig. 2. Simulation results of PD, SW-XPD, HR-PD, and HR-XPD without noise.** (a) Images reconstructed using conventional PD, SW-XPD, HR-PD, and HR-XPD. (b) Axial intensity profiles along the red line in (a), with an enlarged view of the highlighted region shown in the blue box. (c) Box plots of background intensity values extracted from the range indicated by the double-arrow gray line in (b). The central line represents the median, and the whiskers indicate the minimum and maximum values, thus reflecting the full range of background intensities. (d) Estimated resolution improvement obtained from the selected tube cross-section [marked with blue line in (a)] based on the FWHM value. (e) PVL measured from (b), with the dashed line indicating the 3 dB threshold.

As shown in Fig. 2(a, b), PD and HR-PD suffer from strong background and sidelobe artifact (pointed by white arrows), whereas HR-XPD effectively suppresses incoherent interferences while preserving peak signals. Quantitative analysis [Fig. 2(c)] confirms the low background intensity of HR-XPD (mean: –53.22 dB) and SW-XPD (–47.48 dB), markedly lower than PD (–31.43 dB) and HR-PD (–34.05 dB). Resolution assessment further demonstrates that HR-XPD achieves the highest PVL values over 3 dB [Fig. 2(d)]. Consistently, Fig. 2(e) demonstrates that HR-XPD yields resolution improvement of ~1.2-1.4 fold over PD and SW-XPD, confirming its ability to improve spatial resolution while suppressing sidelobe interference.

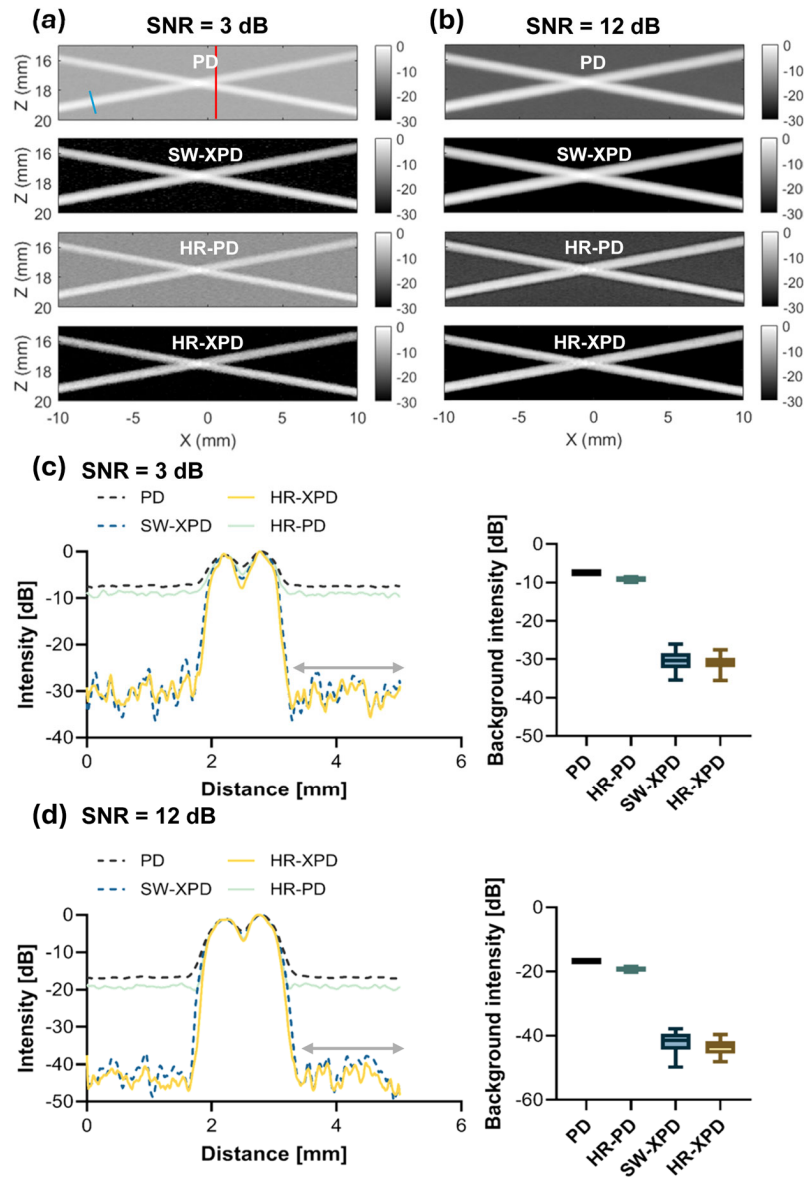

**Fig. 3. Robustness of PD, SW-XPD, HR-PD, and HR-XPD under different SNR conditions.** (a, b) Representative PD images reconstructed with the four methods at SNRs of 3 dB and 12 dB. (c, d) Axial intensity profiles along the red line indicated in (a), showing performance across different SNR levels. The box plots compare background intensity values extracted from the region marked by the double-arrow gray line in the left profile plots. The central line represents the median, and the whiskers indicate the minimum and maximum values, thus reflecting the full range of background intensities.

To investigate the robustness of each method under degraded signal conditions, simulations were conducted at SNR ranging from -2 to 18 dB. Representative Doppler images of PD, SW-XPD, HR-PD, and HR-XPD at SNR levels of 3 dB and 12 dB are shown in Fig. 3(a, b). As the SNR decreases, PD and HR-PD suffer from substantial background noise, while SW-XPD and HR-XPD maintain a high image contrast. The corresponding axial intensity profiles at representative SNR levels are presented in Fig. 3(c, d). For PD and HR-PD, background intensity rises evidently with decreasing SNR, resulting in diminished contrast. By comparison, SW-XPD and HR-XPD effectively preserve vessel visibility and suppress background noise. Notably, HR-XPD consistently achieves low background intensity and clear vessel delineation across all SNR settings, underscoring its robustness to noise.

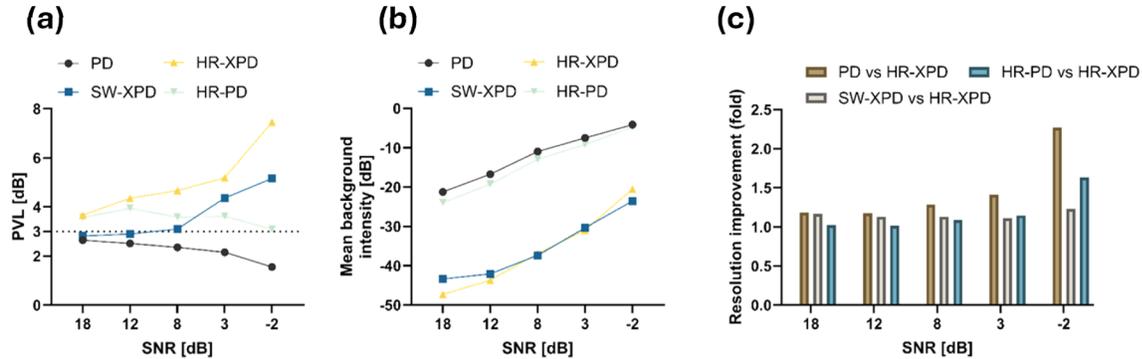

**Fig. 4.** **Quantitative evaluation of PD, SW-XPD, HR-PD, and HR-XPD under different SNR conditions.** (a) PVL calculated from one-dimensional profiles at the locations indicated by the red line in Fig. 3(a). (b) Mean background intensity measured within the background regions, shown in Fig. 3(c, d). (c) Estimated resolution improvement obtained from the selected tube cross-section [marked with blue line in Fig. 3(a)], based on the FWHM value.

As shown in Fig. 4(a), HR-XPD maintains the highest PVL values, consistently above the 3 dB threshold across a wide SNR range, indicating its ability to resolve adjacent flow channels under strong noise interference. Although HR-PD achieves relatively high PVL values (>3 dB) at high SNRs, its performance deteriorates markedly as noise increases. The background intensity results in Fig. 4(b) and the box plots in Fig. 3(c, d) further confirm the effective noise suppression of HR-XPD. Across the entire SNR range, HR-XPD and SW-XPD maintain the lowest background levels, whereas PD and HR-PD exhibit substantially higher background intensity, which increases with decreasing SNR.

The resolution improvement results are summarized in Fig. 4(c). HR-XPD consistently improves resolution compared with PD, HR-PD, and SW-XPD across all SNR levels. The largest gain is observed over conventional PD, especially at low SNR (–2 dB, >2-fold improvement). The incorporation of similarity weighting helped suppress background noise at low SNR thereby maintaining the resolution performance of SW-XPD and HR-XPD with increased noise levels. Compared with HR-PD, HR-XPD still achieves ~1.6-fold enhancement at the SNR level of -2 dB. The above findings validate the effectiveness of HR-XPD in preserving both high contrast and spatial resolution under noise-challenged conditions.

## Kidney transplant

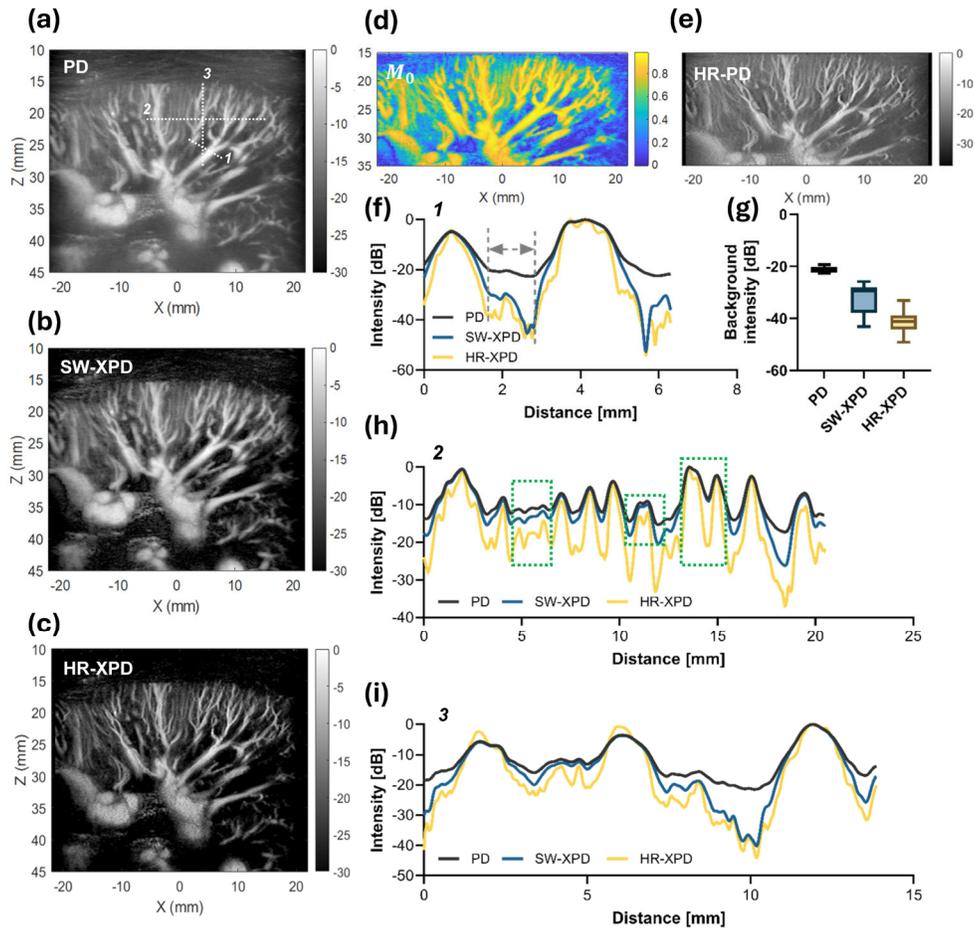

**Fig. 5. Evaluation of PD, SW-XPD, and HR-XPD in a transplanted human kidney.** (a–c) Power Doppler images obtained using PD, SW-XPD, and HR-XPD. (d) Corresponding similarity map $M_0$. (e) HR-PD image. (f, h, i) Normalized intensity profiles along three selected vessel tracks (labeled 1–3 in a). (g) Box plots showing background intensity values extracted from the range denoted by the double-arrow gray line in the profile (f).

Power Doppler imaging was performed on a transplanted human kidney to evaluate the capability of PD, SW-XPD, HR-PD, and HR-XPD in resolving complex vascular structures. As shown in Fig. 5(a–c), conventional PD suffers from high background levels and poor resolvability of small vessels. SW-XPD improves vessel contrast by suppressing background noise and artifacts, whereas HR-XPD provides the sharpest visualization of the vascular network. The corresponding similarity map $M_0$ and HR-PD image are presented in Fig. 5(d, e) further illustrate the contrast enhancement achieved through similarity map weighting in both SW-XPD and HR-XPD.

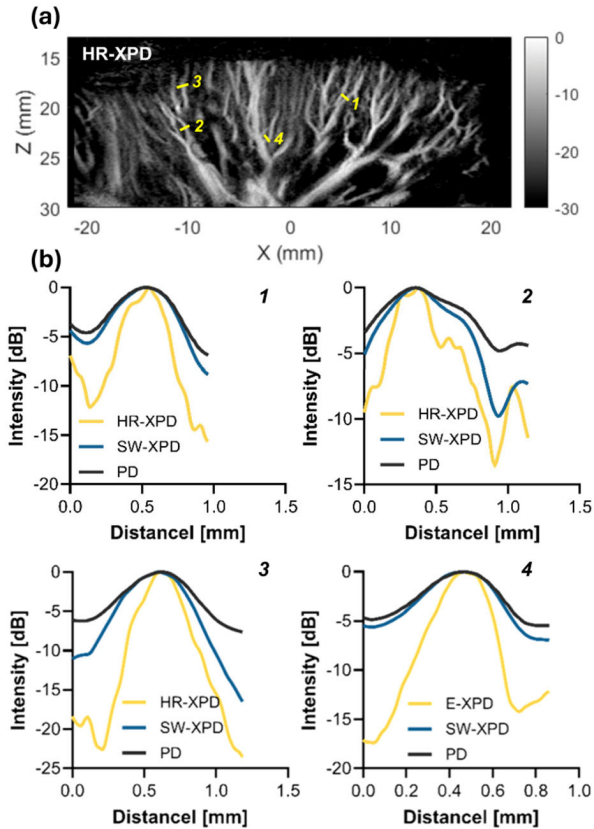
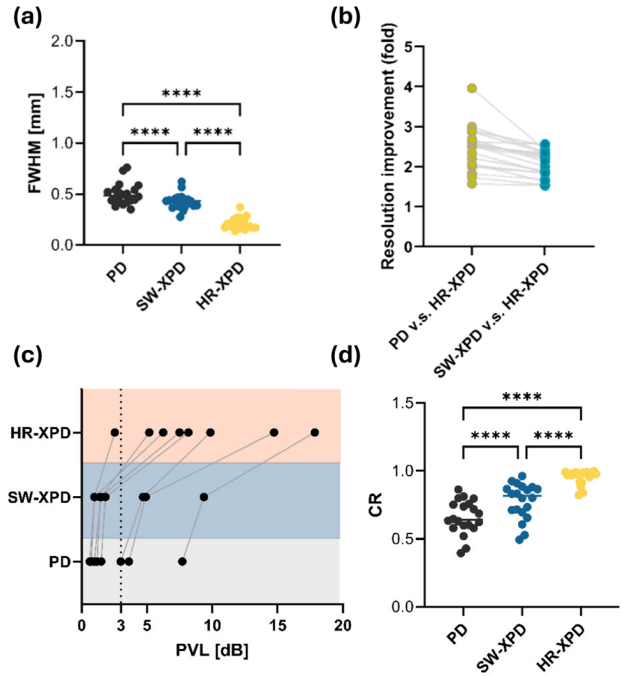

**Fig. 6. FWHM across selected vessel segments in the transplanted kidney.** (a) HR-XPD image. (b) Normalized cross-sectional profiles of representative vessel segments (marked in a).

**Fig. 7. Quantitative analysis of spatial resolution and contrast across selected vessel segments in the transplanted kidney.** (a) FWHM calculated from 20 manually selected vessel cross-sections. (b) Resolution improvement ratios of HR-XPD versus PD and SW-XPD. Each gray line connects the same vessel imaged under different methods. (c) PVL values measured from 8 randomly selected cross-sections, with each line linking same measurement position. (d) CR values of 20 manually selected vessel segments. Statistical comparisons were performed using two-sided paired one-way ANOVA. Significance levels: ****$p < 0.0001$.

Intensity profiles along three selected vessel paths [Fig. 5(f, h, and i)] demonstrate that HR-XPD preserves high vessel-to-background contrast and enables clearer separation between adjacent vessels (marked with green rectangles). Consistently, the box plot in Fig. 5(g) shows substantially reduced background intensity with HR-XPD (mean: –41.17 dB) and SW-XPD (mean: –32.46 dB), compared with PD (mean: –21.25 dB), confirming the superior background suppression of the proposed methods.

Figures 6 and 7 summarizes the quantitative evaluation of imaging performance evaluation of imaging performance across different techniques. FWHM values were measured from 20 manually selected vessel cross-sections, with representative local profiles shown in Fig. 6(a, b). As illustrated in Fig. 7(a), HR-XPD yielded the smallest FWHM values, with paired one-way ANOVA confirming statistical significance ($p < 0.0001$). Resolution improvement was further quantified by the ratio of measured FWHM values [Fig. 7(b)], showing that HR-XPD provided a 2.46-fold enhancement over PD and a 2.08-fold enhancement over SW-XPD. Consistently, PVL analysis from eight typical cross-sections [Fig. 7(c)] demonstrated markedly higher values for HR-XPD, indicating improved separation of adjacent vessels that PD and SW-XPD failed to resolve (< 3 dB). CR results in Fig. 7(d), derived from the same profiles, further confirmed that HR-XPD achieved significantly higher contrast than both SW-XPD and PD (p < 0.0001).

Together, these results demonstrate that HR-XPD substantially improves both spatial resolution and vascular contrast *in vivo*, highlighting its potential utility for clinical assessment of complex renal microvascular structures.

## Healthy human liver

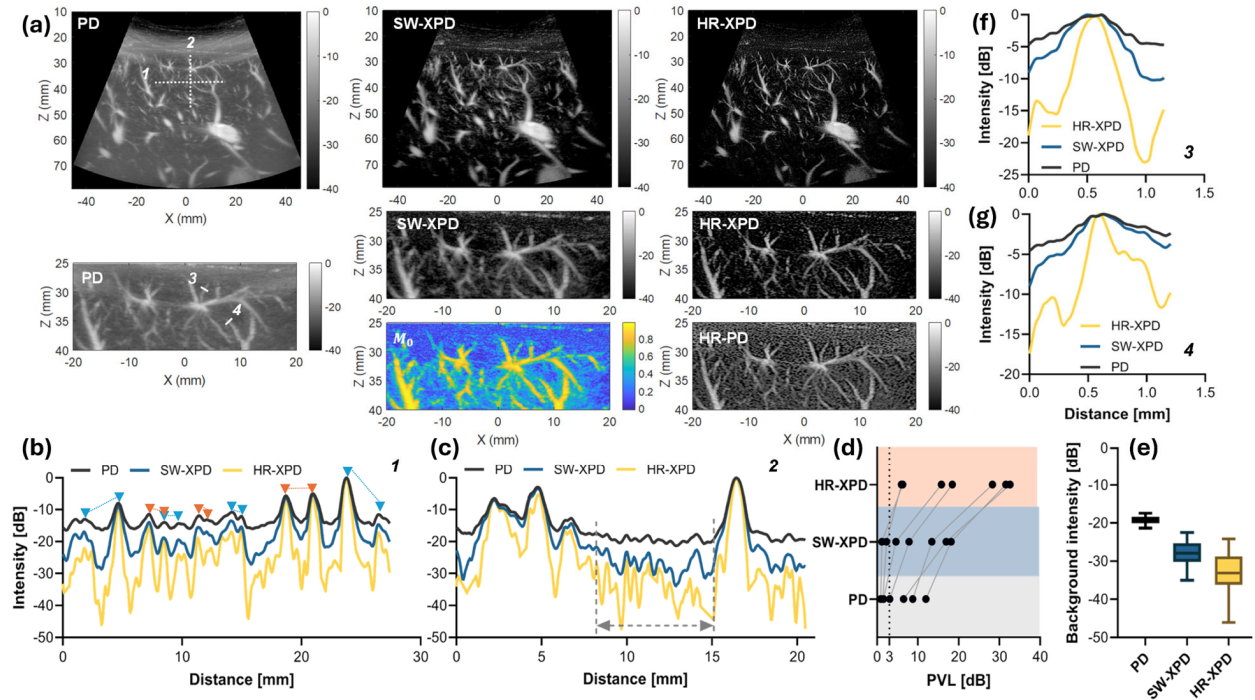

**Fig. 8.** *In vivo* **comparison of healthy human liver data**. (a) Power Doppler images reconstructed using PD, SW-XPD, HR-PD, and HR-XPD, with corresponding zoom-in views and similarity map $M_0$. (b, c) Normalized lateral and axial intensity profiles along white dotted lines 1 and 2 in (a), respectively. (d) PVL values measured from 7 randomly selected cross-sections, with each line linking the same measurement position. The peak pairs used for PVL calculation are color-coded, corresponding to the nearby colored markers in panel (b). (e) Box plots showing background intensity values extracted from the range denoted by the double-arrow gray line in the profile (c). (f, g) Normalized intensity profiles across two vessel profiles, indicated as 3 and 4 in (a).

*In vivo* imaging was also conducted on the liver of a healthy human subject, and representative images are shown in Fig. 8 (a). Compared with conventional PD and HR-PD, which exhibit elevated background and reduced vessel contrast, SW-XPD enhances vessel visibility, while HR-XPD provides further improvement with higher spatial resolution. Lateral and axial intensity profiles along the white dotting lines [Fig. 8 (b, c)] demonstrate that HR-XPD yields deeper valleys and sharper peaks, indicating improved vessel separation and effective background suppression in both directions. The corresponding background intensities [Fig. 8 (e)] further confirm this advantage, showing that HR-XPD (mean: –32.95 dB) and SW-XPD (–27.90 dB) achieve substantially lower levels than PD (–19.28 dB).

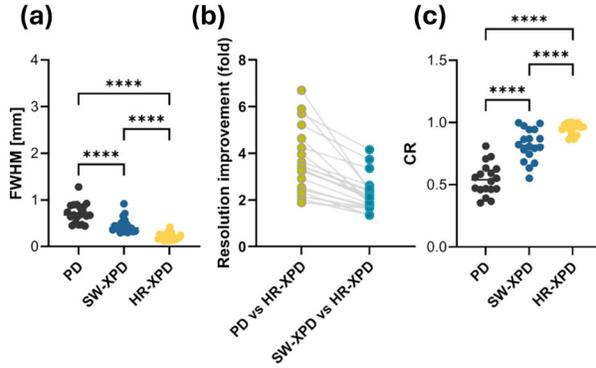

**Fig. 9. Quantitative comparison of healthy human liver data. (a) FWHM analysis across 20 cross-sectional profiles of manually selected vessels.** (b) Resolution improvement across vessel pairs. Each gray line connects matched vessels. (c) CR values are computed from the cross-sections of 20 manually selected vessels. Statistical analysis was performed using two-sided paired one-way ANOVA tests. Significance levels: ****p<0.0001.

Spatial resolution was quantified using FWHM measurements and PVL values. As shown in Fig. 9 (a), HR-XPD significantly reduces FWHM compared to both PD (p<0.0001) and SW-XPD (p<0.0001). The FWHM ratios across vessel segments [Fig. 9(b)] indicate that HR-XPD achieves, on average, a 3.65-fold improvement over PD and a 2.25-fold improvement over SW-XPD, reflecting substantial gains in resolving capability. Consistent with these findings, Fig. 8(d) shows that HR-XPD yields higher PVL values (mean: 19.85 dB) than PD (4.87 dB) and SW-XPD (9.21 dB), enabling separation of 3 adjacent vessel pairs that conventional PD fail to resolve (< 3 dB). Finally, CR comparison [Fig. 9(c)] demonstrates that HR-XPD significantly enhances vessel-to-background contrast, achieving higher values than both SW-XPD and PD (p<0.0001).

The above results confirm that HR-XPD provides superior *in vivo* performance by simultaneously enhancing spatial resolution and suppressing background noise, enabling clearer visualization of hepatic microvascular structures.

## Pig kidney

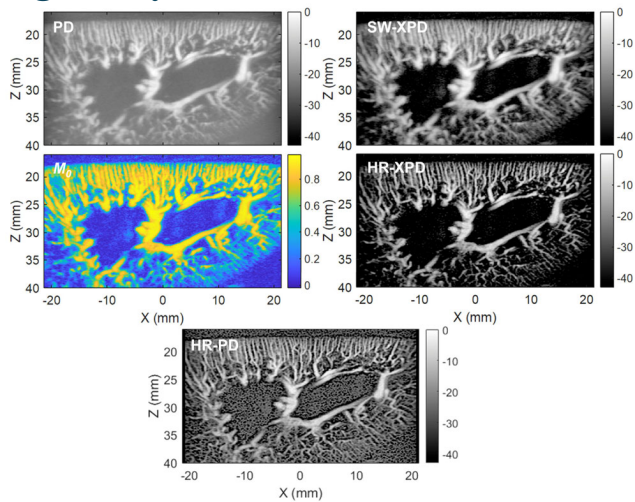

**Fig. 10.** *In vivo* **imaging of pig kidney data. PD, SW-XPD, HR-XPD, HR-PD images, and similarity map ($M_0$) obtained using an ensemble of 300 frames.**

Figure 10 demonstrates the performance of PD, SW-XPD, HR-PD, and HR-XPD in pig kidney imaging. Reconstructed images show that conventional PD suffers from high background noise and poor visibility of fine vascular structures, whereas SW-XPD suppresses background interference and improves vessel contrast. The proposed HR-XPD further enhances vessel sharpness and delineation of small vascular branches, even when using only 0.3 s of data acquisition.

The lateral intensity profiles in Fig. 11(b) corroborate these findings: HR-XPD produces substantially narrower vessel width compared with PD and SW-XPD, resulting in improved PVL values [Fig. 11(c)]. Across paired comparisons of 11 sampling vessel pairs [Fig. 11(b)], SW-XPD provided limited improvement in PVL (mean: 5.10 dB) compared to PD (3.69 dB), with 64% (7/11) of measurements exceeding the 3 dB threshold (PD: 5/11). In contrast, HR-XPD markedly improved the PVL (mean: 16.49 dB) with all the measurements surpassing 3 dB criterion, highlighting its ability to resolve closely spaced vascular structures.

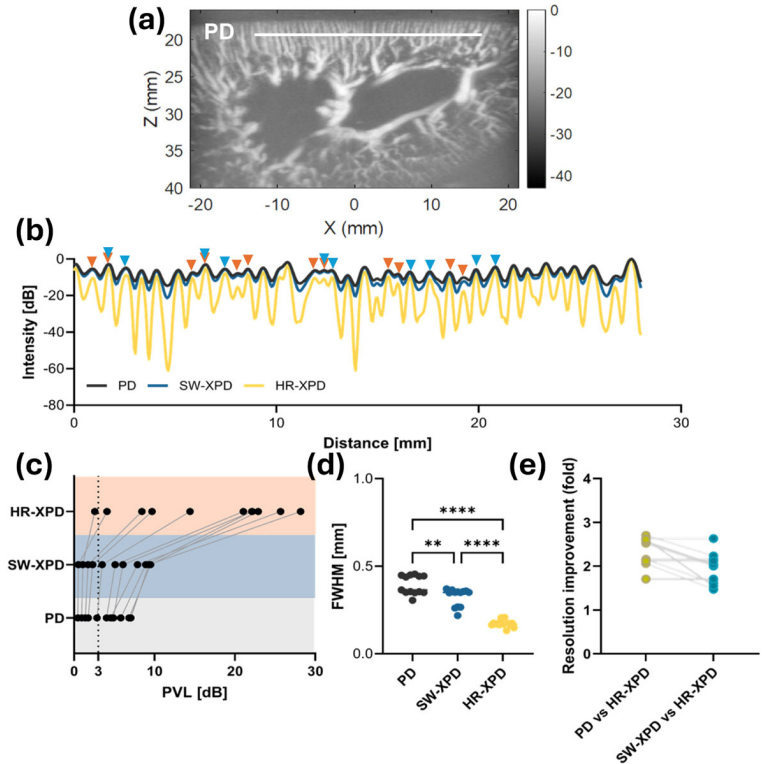

**Fig. 11. Comparison of pig kidney data.** (a) HR-XPD. (b) Normalized lateral intensity profiles along the white line in (a). (c) PVL values measured from 11 randomly selected cross-sections of two adjacent vessels, with each line linking the same measurement position. The peak pairs used for PVL calculation are color-coded, corresponding to the nearby colored markers in panel (b). (d) FWHM calculated from 13 manually selected vessel cross-sections. (e) Resolution improvement ratios of HR-XPD versus PD and SW-XPD. Each gray line connects the same vessel imaged under different methods. Statistical comparisons were performed using two-sided paired one-way ANOVA. Significance levels: **$p < 0.01$, ****$p < 0.0001$.

FWHM analysis [Fig. 11(d)] further demonstrated that HR-XPD consistently achieves the smallest FWHM values across all sampled vessels, with significant differences compared to SW-XPD ($p < 0.01$) and PD ($p < 0.0001$). As summarized in Fig. 11(e), HR-XPD provides ~1.5–2.8-fold resolution improvement relative to PD and SW-XPD. Together, these results show that HR-XPD substantially enhances both vascular contrast and spatial resolution compared with conventional PD and SW-XPD.

## Discussions

In this study, we propose and evaluate a high-resolution power Doppler method, denoted as HR-

XPD, to improve microvascular imaging under diverse experimental and clinical conditions. Compared with conventional PD, SW-XPD, and HR-PD, HR-XPD consistently achieved superior spatial resolution and contrast. *In vivo* experiments in healthy human liver, human kidney transplants, and pig kidney demonstrated significantly suppressed background, with resolution improved by approximately 2–3 fold over PD and SW-XPD. These findings indicate that HR-XPD enables detailed visualization of microvascular morphology and blood volume without the need for contrast agents.

The superior performance of HR-XPD arises from the integration of radiality measurements, which enhance spatial resolution, with similarity weighting and cross-correlation analysis, which mitigate the noise sensitivity inherent to radiality evaluation. Figures 3, 8, and 10 illustrate that radiality enhancement alone (HR-PD) improves resolution but suffers from strong noise interference, confirming the susceptibility of radiality weighting to noise. By further incorporating similarity weighting and cross-correlation analysis, the incoherent noise and background interference can be effectively suppressed in HR-XPD images, while maintaining the resolution gain. This highlights its potential for clinical applications that require accurate visualization and quantification of microvascular networks.

In this study, high-resolution vascular details were clearly depicted within a short acquisition time of only 0.3 s-1.2 s without the use of contrast agents. From a practical perspective, achieving high-quality vascular imaging within short acquisition times is essential for capturing transient hemodynamic changes, reducing motion artifacts, and improving acquisition efficiency. In this study, detailed vascular morphology in pig kidney and transplanted human kidney was clearly depicted within only 300 frames (0.3-0.6 s of data acquisition). This demonstrates that most vascular structures can be revealed with limited data volume. Such capabilities are particularly valuable for functional ultrasound (fUS), where high spatial resolution improves the accuracy of activation mapping and short acquisition times enhance temporal resolution for detecting transient, stimulus-induced blood flow changes.

In the simulation studies, spatial resolution was strongly affected by noise, showing substantial degradation under low-SNR conditions. The incorporation of similarity weighting helped suppress background noise at low SNR, thereby maintaining the resolution performance of SW-XPD and HR-XPD even when noise levels increased. Nevertheless, it should be noted that in our simulation setup the axial resolution was very high, which limited the observable resolution improvement achieved by HR-XPD.

The proposed framework may also offer a potential pathway for extending radiality-based enhancement to SRUS imaging. While radiality-based processing has been demonstrated in CEUS imaging, its super-resolution feasibility has so far been primarily validated in rabbit or mouse models [47, 48]. Its performance and robustness under more challenging conditions, such as varying imaging depths, attenuation and artifacts, remain to be further investigated. Combining CEUS imaging with cross-correlation and similarity map enhancement could mitigate performance degradation in low-SNR scenarios, providing a more robust solution for clinical microvascular imaging. This possibility remains to be explored in future studies.

Despite its advantages, the proposed method has several limitations. As HR-XPD is inherent relying on the blood flow Doppler signal without the need for contrast agent, it is expected to be less sensitive to the smallest vessels compared to microbubble-tracking based ultrasound localization microscopy (ULM) [17, 56]. As the most widely studied SRUS imaging technique, ULM reconstructs highly detailed vascular maps beyond the diffraction limit by tracking sparsely distributed microbubbles over extended acquisition periods [49, 56-59]. However, ULM requires longer acquisition times and stable microbubble concentrations [60, 61], which complicates clinical deployment [28], particularly when bolus injection protocols are adopted in many workflows [28]. In contrast, HR-XPD offers a higher temporal resolution, making HR-XPD particularly attracting for future real-time and specific clinical applications. While this study primarily focused on vascular resolvability and contrast enhancement, HR-XPD may be further extended by integrating with color Doppler to provide flow velocity information in the future.

## Conclusion

By integrating radiality weighting with spatiotemporal cross-correlation and similarity evaluation, HR-XPD simultaneously enhances spatial resolution and image contrast. In multiple *in vivo* studies, HR-XPD consistently demonstrated higher contrast and lower background levels compared with conventional PD and HR-PD. Resolution assessments revealed an approximately 2–3-fold improvement over PD and SW-XPD, enabling detailed visualization of microvascular morphology and blood volume without the need for contrast agents. Taken together, HR-XPD represents a promising approach for high-resolution microvascular imaging, holding strong potential for broad clinical translation.

## Acknowledgement

The study was partially supported by the National Institute of Diabetes and Digestive and Kidney Diseases under award numbers of R01DK129205 and R01DK138998. The content is solely the responsibility of the authors and does not necessarily represent the official views of the National Institutes of Health.

## Conflict of interest

The Mayo Clinic and some of the authors (J.Y., J.Z., S.C. and C.H.) have pending patent applications related to the technologies referenced in this publication. Additionally, the Mayo Clinic and some of the authors (J.Z., S.C. and C.H.) have a potential financial interest related to the part of the technologies used in this study.

# Supplementary information

TABLE S1
PARMETER SETTINGS OF *IN VIVO* DATA

|  | *In vivo* kidney transplant | *In vivo* healthy human liver | *In vivo* pig kidney |
|---|---|---|---|
| Transducer | Linear array transducer GE 9L-D | Curved array transducer C1-6-D | Linear array transducer GE 9L-D |
| Transmit center frequency (MHz) | 5.28 | 4.46 | 5.28 |
| Number of elements | 192 | 192 | 192 |
| Pitch (mm) | 0.23 | 0.35 | 0.23 |
| Transmission type | Plane wave | Diverging wave | Plane wave |
| Steering Angles | −9° to +9° with 2° increment | −4.5° to +4.5° with 1° increment | −9° to +9° with 2° increment |
| Post-compounded frame rate (Hz) | 500 | 500 | 1,000 |
| Transmit one-sided voltage (V) | 50 | 50 | 50 |
| Total number of frames | 300 | 600 | 300 |
| Pixel size of IQ (mm × mm, axial × lateral) | 0.06 × 0.12 | 0.17 × 0.17 | 0.06 × 0.12 |